\documentclass{rrparticle}
\usepackage{graphicx}

\newcommand{\miktex}{\hbox{Mik\kern-.15em\TeX}}

\title{Video analysis of the damped oscillations of Pohl's pendulum} 
\author[1]{Ivan Z. Stefanov}
\author[2]{Nikolay Denev}
\author[3]{Sava Donkov}
\affil[1]{Department of Applied Physics, Technical university of Sofia, \authorcr 8, Snt. Kliment Ohridski Blvd.,   1000 Sofia, Bulgaria, \authorcr Email:{ \em izhivkov@tu-sofia.bg}}
\affil[2]{Department of Applied Physics, Technical university of Sofia, \authorcr 8, Snt. Kliment Ohridski Blvd.,   1000 Sofia, Bulgaria, \authorcr Email:{\em ndenevtph@tu-sofia.bg}}
\affil[3]{Institute of Astronomy and NAO, Bulgarian Academy of Sciences, 72 Tsarigradsko Chausee Blvd., 1784 Sofia, Bulgaria, \authorcr Email:{\em sddonkov@astro.bas.bg}}
\DeclareUnicodeCharacter{2212}{-}
\begin{document}
	\maketitle
\begin{abstract}
		In this paper problems that arose with the introduction of distance learning in physics at the Technical University of Sofia due to the COVID-19 pandemic and the imposition of video recording of laboratory exercises are indicated. It was found that the video for the “Damped Mechanical Oscillations” exercise provides enough information for a more detailed and in-depth analysis of the studied phenomenon compared to the standard way of capturing the data. The Video Editor program was used to view the video frame by frame and statistical processing - non-linear regression - was performed with the recorded data. The laboratory results are compared with the theoretical function, the parameters of which are optimized as a result of the specified processing. A theoretical model of the damped oscillation is described and the dependence of the damping coefficient on the current through the electromagnetic brake is theoretically investigated.
\end{abstract}
\section{Introduction}
The COVID-19 pandemic has inflicted a number of changes and limitations upon our lives. This inevitably happened in the field of secondary and higher education. Due to the need to work from home, platforms such as Moodle, Discord, Microsoft Teams, etc. had to be used to facilitate distance learning. At the same time, other measures had to be taken to improve the quality of teaching under these conditions. The Department of Applied Physics at the Faculty of Applied Mathematics and Informatics of the Technical University of Sofia (TU - Sofia) decided at the beginning of the pandemic to prepare video demonstrations of the laboratory exercises in physics and brief descriptions. These videos were used by the students to get a better idea of the exercise and the equipment used. 
As it can be seen in \cite{DataLogger,FromYoutube,ArtificialVideo,VideoChain,ArduinoTrackerRRP,LaserRRP}, for example, video analysis of laboratory exercises has been gaining  popularity lately.

We later found that some of the videos contained detailed information concerning the measured data. This gave rise to the idea to improve one of the exercises, using footage from its video in order to capture laboratory data with the help of a special program. This program is Windows 10's built-in "Video Editor". With its help, the frames can be viewed one by one and data about processes that take place over time can be recorded. Here the time step is 0.033~s. The motion we have considered is damped oscillation with a period of about 1.8~s. The part of the video we use has a duration of 4.5~s, i.e. it includes 2.5 periods and data of about 130 frames. The presence of a large number of data allows for statistical processing, in our case we use nonlinear regression.

\section{Experimental setup}
\begin{figure}[ht!]
	\centering
	\includegraphics[width=1.0\textwidth,keepaspectratio]{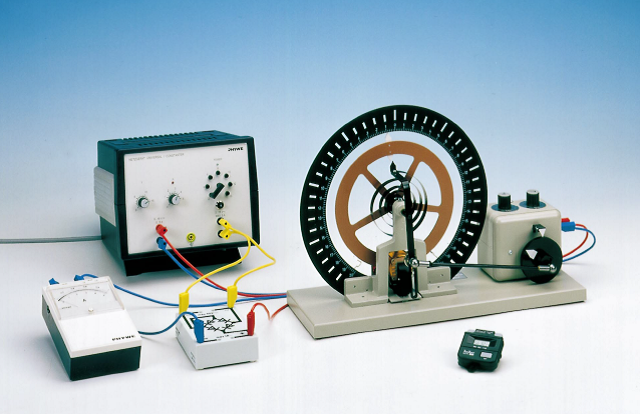}
	\caption{Photo of the experimental setup. (Taken from the original instruction manual of PHYWE \cite{PHYWE}.)}\label{Setup}
\end{figure}

The name of the exercise we focused on is "Damped mechanical oscillations". The laboratory set is produced by Leybold Didactic\footnote{An almost identical setup is available also by PHYWE.} and represents Pohl’s Pendulum, Figure~\ref{Setup}.
In addition several PHYWE modules were used in the laboratory set.

Pohl's pendulum is a torsion pendulum, composed of a brass disk attached to a coil spring. As it deviates from equilibrium and relaxes, the disk oscillates about a horizontal axis passing through its center. The oscillation is damped and the damping factor can be adjusted. The lower part of the disk passes between the poles of an electromagnet and Foucault currents are induced in it. They interact with the field of the electromagnet as a result of which a restoring momentum acts on the disk, i.e. the system works as an additional system for the increasing of the damping coefficient -- an electromagnetic brake. By changing the current through the electromagnet, we change this factor.
To measure the deviation of the disk from its equilibrium position, an index finger is attached at its upper rim. There is also a metal strip with a scale in relative divisions which is concentric to the disk. At the beginning, the disk is rotated until the pointer indicates the desired initial deviation, held and released at the same time as the timer is started. The disk performs damped oscillations, i.e. its maximum deviation (the amplitude) decreases with time by exponential law.
The standard method of performing the exercise is to measure the oscillation amplitude of the pendulum at the end of each period, starting with a given initial deviation at the starting point in time. Then a graph of the dependence of the amplitude on time is plotted.
\section{Model}
According to \cite{PHYWE} the damped oscillations of Pohl's pendulum are described by the following differential equation
\begin{equation}\label{ODE}
	\ddot{\varphi}+2\gamma\dot{\varphi}+\omega_0^2\varphi=0,
\end{equation}
where $\varphi$ is the angular displacement. The two parameters which determine the motion are the natural angular frequency $\omega_0^2$ and the damping constant $\gamma$. The former characterizes the undamped system and is determined by   
$$\omega_0^2 = \frac{D^0}{I} ~,$$
where $D^0$ describes the spring's torque per unit angle and $I$ is the moment of inertia of the disk. For the latter parameter the following formula is given
$$\gamma = \frac{C}{2I} ~,$$
where $C$ is a factor of proportionality in the equation for torque $M = - C \dot{\varphi}$ produced from the eddy current brake (where $\dot{\varphi}$ is the angle velocity of the pendulum). As it is claimed in \cite{PHYWE} the factor $C$ depends on the current $I_{\rm B}$ through the eddy current brake. Let us asses how this dependence looks like.

We know that the torque $M$ results from the interaction between the magnetic fields of the electromagnet and these created by the Foucault currents induced when the disk passes between poles of the electromagnet. Let us imagine a Foucault current produced in the volume element of the disk (flowing parallel to the disk surface) at the moment when this volume crosses the magnetic field of electromagnet. Let the characteristic size of this volume element be $\Delta l$. The induced Foucault current represents a magnetic dipole with a magnetic dipole moment $m \sim I_{\rm i} S$, where $I_{\rm i}$ is the strength of Foucault current, and $S \sim \Delta l ^2$ is the surface of this current. From the Faraday law $I_{\rm i} = - (1/r) \Delta\Phi_{\rm B}/\Delta t$, where $r$ is the electric resistivity of the current contour, $\Delta\Phi_{\rm B} \sim B S$ is the change of magnetic flow through the contour ($B \sim I_{\rm B}$ is the magnetic field of electromagnet), and $\Delta t \sim (\Delta l/R) \dot{\varphi}^{-1}$ is a small time for the process in regard ($R$ is a radius at which the considered volume element resides). In a mechanical sense, the torque $M$ is a result of the change of potential energy $\Delta U$ of the magnetic dipole when it crosses the border of the magnetic field $B$ of electromagnet. So $M \sim R \Delta U / \Delta l$, where $\Delta U / \Delta l$ is the gradient of the potential energy and $\Delta U \sim mB \sim I_{\rm i}SB$. Finally, combining all the above relations, one obtains:
$$ M \sim R \frac{\Delta U}{\Delta l} \sim R \frac{I_{\rm i}SB}{\Delta l} \sim \frac{R}{r} \frac{S^2 B^2}{\Delta l \Delta t} \sim \frac{1}{r} \bigg(\frac{RS}{\Delta l}\bigg)^2 B^2 \dot{\varphi} \sim I_{\rm B}^2 \dot{\varphi} ~. $$
Hence we obtain that the damping index $\gamma$ is proportional to the $I_{\rm B}^2$, with an accuracy up to a geometrical factor and constants characterizing electrical and magnetic properties of the material of the disk.  

The solution of equation (\ref{ODE}) is \cite{PHYWE}
\begin{equation}\label{model}
	\varphi(t)=\varphi_0e^{-\gamma t}\cos(\omega t + \alpha),
\end{equation}
where $\alpha$ is the initial phase and $\omega$ is angular frequency of the oscillations 
$$\omega=\sqrt{\omega_0^2+\gamma^2}.$$
\section{Video analysis of the damped oscillations}
Here the idea is to measure the deviation of the pendulum from the equilibrium position in steps of 0.033~s\footnote{On several occasions two adjacent frames were identical. In these cases we chose to keep only one of them and neglect the other. } by analyzing the video frame by frame and thus to track the movement within about 5~s. As it was mentioned above, the period of the damped oscillations is about 1.8~s, so the video gives information about 2.5 periods and contains 126 frames. 

\begin{figure}[ht!]
	\centering
	\includegraphics[width=1.0\textwidth,keepaspectratio]{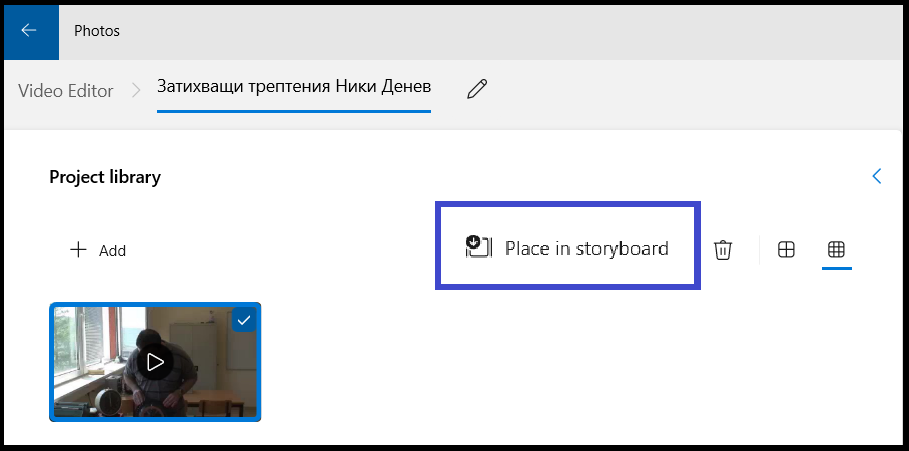}
	\caption{Photo of the interface of Video Editor.}\label{Place in storybard}
\end{figure}

\begin{figure}[ht!]
	\centering
	\includegraphics[width=1.0\textwidth,keepaspectratio]{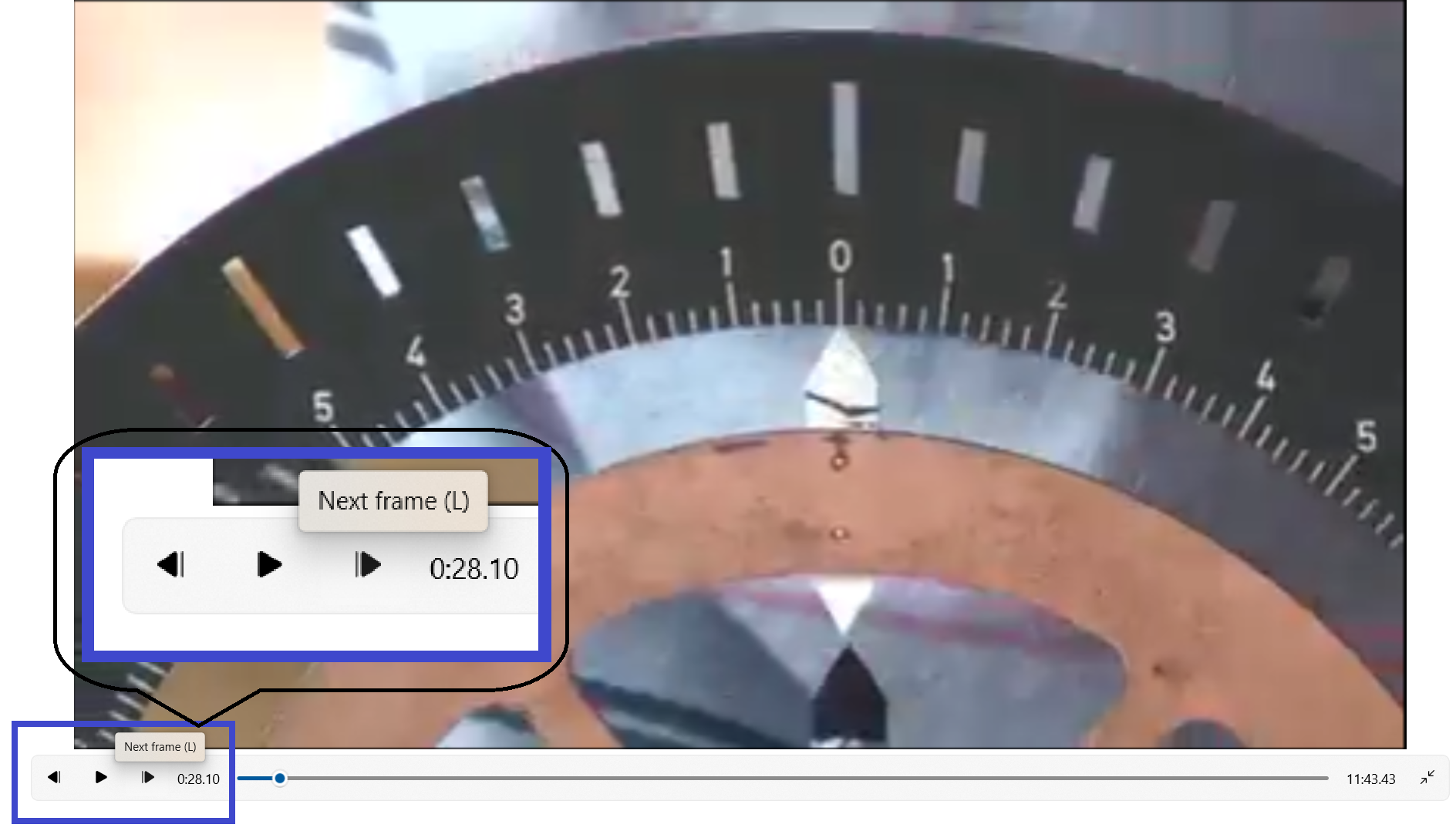}
	\caption{Photo of the  buttons in Video Editor used for scrolling through the frames.}\label{Buttons}
\end{figure}

In order to analyze the video frame by frame we use the Windows 10's build-in Video Editor. Before being able to manipulate a video one has to open it with the program and then to place it in the storyboard using the button framed in blue on Figure~\ref{Place in storybard}. In order to scroll through the frames one can use either the interface buttons, shown on Figure~\ref{Buttons}, or use the letter "L" on their keyboard as a short key.

The data obtained from the video is entered into Excel and processed statistically. The result is information about the amplitude, the damping coefficient, the angular frequency and the initial phase of the oscillations. The logarithmic decrement of damping and period can also be calculated.

The advantages of this method of analysis by video recording are that the information obtained here is much richer and more accurate than the standard method of conducting the exercise.
\section{Reduction of data}
\begin{figure}[ht!]
	\centering
	\includegraphics[width=1.0\textwidth,keepaspectratio]{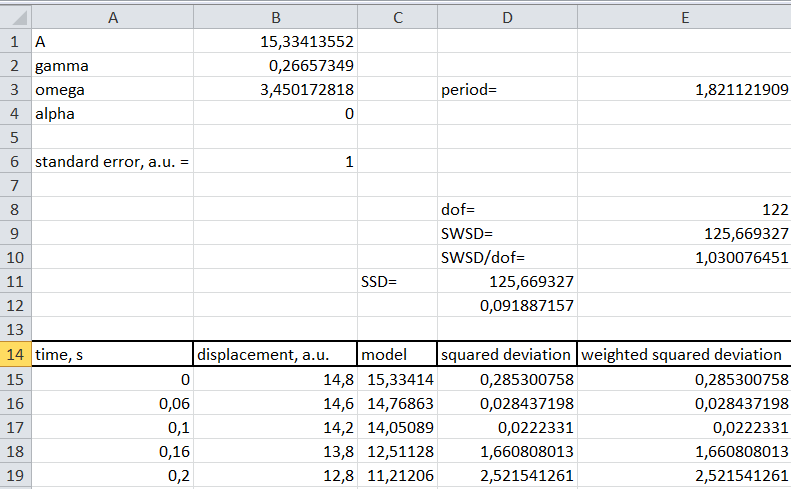}
	\caption{The excel (top part) table containing the data, the model parameters and the merit function -- SSD}\label{xlsTable}
\end{figure}

\begin{figure}[ht!]
	\centering
	\includegraphics[width=1.0\textwidth,keepaspectratio]{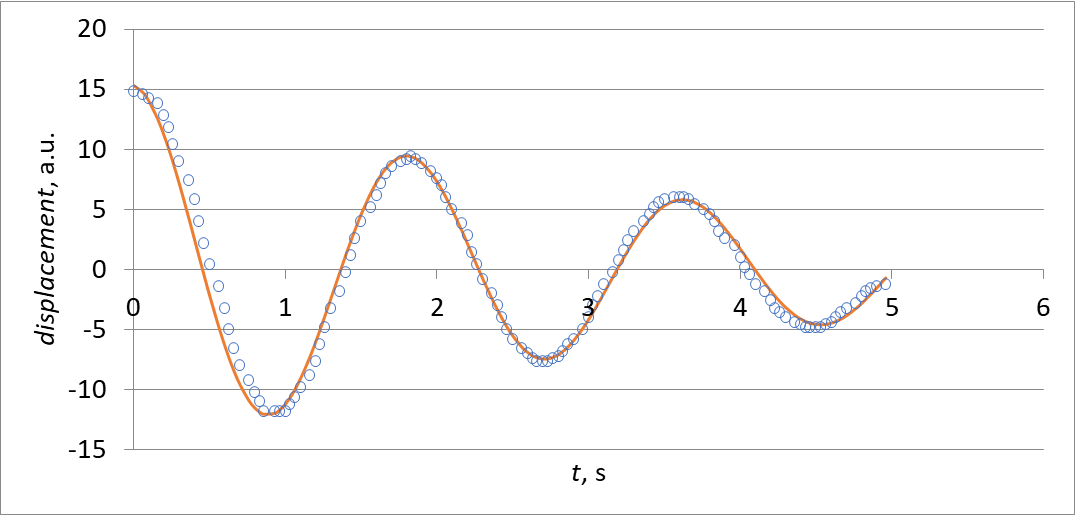}
	\caption{Experimental data designated by circles, and model curve -- solid line.}\label{data}
\end{figure}

We use Microsoft Excel to process the experimental data. The data is arranged in the table shown on Figure~\ref{xlsTable}. 
The first column contains the time values with a step of 0.033~seconds, and the second column shows the readings of the pendulum deviations from the equilibrium position measured in arbitrary units. The experimental points are designated by circles on Figure~\ref{data}.
The values obtained with the help of the model function (\ref{model}) are recorded in the third column of the Excel table. The fourth column contains the squared differences between the measured values and those calculated with model. Their sum is placed in cell D11, as can be seen in Figure~\ref{xlsTable}.

\begin{figure}[ht!]
	\centering
	\includegraphics[width=1.0\textwidth,keepaspectratio]{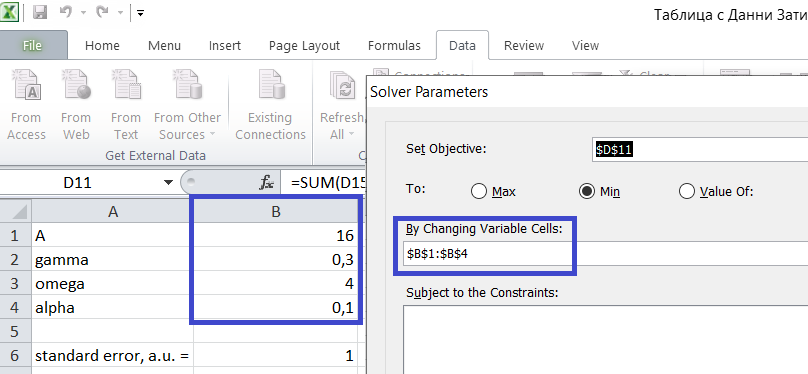}
	\caption{The dialog box of Solver.}\label{xlsSolverDialog}
\end{figure}

The model function has four free parameters: amplitude, damping index, circular frequency and initial phase. Their values are entered in cells B1, B2, B3 and B4, respectively. When constructing the model curve indicated by the orange color on Figure~\ref{data}, we enter approximate values of these parameters obtained from a rough estimate. (See for example the values surrounded by blue in Table \ref{xlsSolverDialog}.)
Estimates of the optimal values of the model parameters can be obtained by the method of nonlinear regression. An explanatory video for the application of the nonlinear regression method in Microsoft Excel can be seen in the YouTube channel ENGR 313 - Circuits and Instrumentation \cite{NonLinearFitYouTube}. 

In short, this method seeks the minimum of the objective function, namely the sum of the squared differences\footnote{SSD is the term used in \cite{NonlinearRegressionEngineers} and in the instructive video in YouTube for the application of the nonlinear regression method in Microsoft Excel \cite{NonLinearFitYouTube}. This quantity is also termed SSE or sum of the squared errors \cite{StudentsGuide}. } (SSD) between the measured values  and those calculated using the model for the deviation of the pendulum from its equilibrium position at times $t_{\rm i}$.

To minimize the objective function SSD, i.e. the value stored in cell D11, we use the Solver function, which is located in the Data menu. If this function is not visible, it must be enabled from the Properties menu. Select the cell whose value we want to optimize and then call Solver from the Data menu. The dialog box shown in Figure~\ref{xlsSolverDialog} appears. 
In this dialog box, we must note that we want to minimize the value of a given cell, specify its position, and specify which cell values should be varied. These cells are surrounded by blue, in Figure~\ref{xlsSolverDialog}.

The optimal values of the parameters that we obtain are (See also Figure~\ref{xlsTable}.):
\begin{eqnarray}
	A=15.3\\
	\gamma=0.27\\  
	\omega=3.45 \,\,{\rm s^{-1}}\\  
	\alpha=0 \,\,{\rm rad}.   
\end{eqnarray}
The period that we obtain then is $T=1.82\,\,{\rm s}$, which is in agreement with the value cited in \cite{PHYWE}. 

As an estimate of the quality of the approximation we can use the chi-square test \cite{NumericalRecipies, PeterScott, Bevington}. The chi-square variable differs from the SSD in that the squared differences between the measured values and those calculated using the model are weighted by experimental error, $w_{\rm i}=1/\sigma_{\rm i}^2$, or
\begin{equation}
\chi^2=\sum_{i=1}^{N} \frac{\left(\varphi(t_{\rm i})-\varphi_{\rm i}\right)^2}{\sigma_{\rm i}^2},
\end{equation}
where $\varphi_{\rm i}$ are the measured deviation of the disk, while the values $\varphi(t_{\rm i})$ are those calculated with model (\ref{model}).
The sum of the weighted squares of the deviations is entered in cell E9, from the table in Figure~\ref{xlsTable}. The degrees of freedom (dof) are obtained by subtracting the number of parameters from the number of experimental data point. We can consider that the approximation is good if the reduced the chi-square, i.e. the chi-square divided by the dof, is close to 1.
As it can be seen from Figure~\ref{Setup}, it is normal to take one of the small divisions of the arc as an estimate of the experimental error. With such a choice, however, we obtain a large value of the reduced chi-quadra, i. e. the quality of the approximation is poor. There is no good agreement between the model and the experimental data. A better estimate of the quality of the approximation can be obtained, if the estimate of the experimental error is increased. If we choose, for example, that it is equal to one of the large divisions of the scale, then the reduced chi-square, given in E10 on Figure~\ref{xlsTable}, is 1.03 -- definitely acceptable.
\section{Conclusion}
In the presented work, we use the Video Editor program for frame-by-frame analysis of the video of the exercise "Damped Mechanical Oscillations ". We process the captured data statistically by non-linear regression. The experimental results were compared with the model oscillation curve, as a result of which the process parameters were optimized. A theoretical model of damped oscillation is presented and the dependence of the damping coefficient on the current through the electromagnetic brake is theoretically investigated.
We hope that this work will support the teaching of the material on damped mechanical oscillations and improve students' understanding of this phenomenon, while sharpening their attention to some physical processes involved in the realization of the exercise.

\begin{acknowledgement}
The author would like to thank assoc. prof. Stefan Nitsolov for providing the high definition video camera and assoc. prof. Chavdar Hardalov for shooting the video.
\end{acknowledgement}

\end{document}